%% file: sample-sigconf.tex
\documentclass[sigconf]{acmart}
\usepackage{color}

\usepackage{textcomp}
\usepackage{stfloats}
\usepackage{url}
\usepackage{verbatim}
\usepackage{amsmath}
\usepackage{amsthm}
\usepackage{subfigure}
\usepackage{makecell}
\usepackage{multirow}
\usepackage{multicol}
\usepackage{enumitem}

\usepackage{times}

\usepackage{algorithm}
\usepackage{algorithmic}
\usepackage[inkscapelatex=false]{svg}
\AtBeginDocument{%
  }

\setcopyright{acmlicensed}
\copyrightyear{2024}
\acmYear{2024}
\acmDOI{XXXXXXX.XXXXXXX}

\acmConference[Conference acronym 'XX]{Make sure to enter the correct
  conference title from your rights confirmation emai}{June 03--05,
  2018}{Woodstock, NY}
\acmISBN{978-1-4503-XXXX-X/18/06}

\begin{document}

\title{MENTOR: Multi-level Self-supervised Learning for Multimodal Recommendation}


\author{Jinfeng Xu}
\affiliation{%
  \institution{The University of Hong Kong}
  \city{Hong Kong}
  \country{China}}
\email{jinfeng@connect.hku.hk}

\author{Zheyu Chen}
\affiliation{%
  \institution{The Hong Kong Polytechnic University}
  \city{Hong Kong}
  \country{China}}
\email{zheyu.chen@connect.polyu.hk}
 
\author{Shuo Yang}
\affiliation{%
  \institution{The University of Hong Kong}
  \city{Hong Kong}
  \country{China}}
\email{shuoyang.ee@gmail.com}

\author{Jinze Li}
\affiliation{%
  \institution{The University of Hong Kong}
  \city{Hong Kong}
  \country{China}}
\email{lijinze-hku@connect.hku.hk}

\author{Hewei Wang}
\affiliation{%
    \institution{Carnegie Mellon University}  
    \city{Pittsburgh, PA}   
    \country{USA}}
\email{heweiw@andrew.cmu.edu}

\author{Edith C.-H. Ngai}
\authornote{*Corresponding authors}
\affiliation{%
  \institution{The University of Hong Kong}
  \city{Hong Kong}
  \country{China}}
\email{chngai@eee.hku.hk}

\renewcommand{\shortauthors}{Xu et al.}

\begin{abstract}
With the increasing multimedia information, multimodal recommendation has received extensive attention. It utilizes multimodal information to alleviate the data sparsity problem in recommendation systems, thus improving recommendation accuracy. However, the reliance on labeled data severely limits the performance of multimodal recommendation models. Recently, self-supervised learning has been used in multimodal recommendations to mitigate the label sparsity problem. Nevertheless, the state-of-the-art methods cannot avoid the modality noise when aligning multimodal information due to the large differences in the distributions of different modalities. To this end, we propose a \underline{M}ulti-level s\underline{E}lf-supervised lear\underline{N}ing for mul\underline{T}im\underline{O}dal \underline{R}ecommendation (MENTOR) method to address the label sparsity problem and the modality alignment problem. Specifically, MENTOR first enhances the specific features of each modality using the graph convolutional network (GCN) and fuses the visual and textual modalities. It then enhances the item representation via the item semantic graph for all modalities, including the fused modality. Then, it introduces two multilevel self-supervised tasks: the multilevel cross-modal alignment task and the general feature enhancement task. The multilevel cross-modal alignment task aligns each modality under the guidance of the ID embedding from multiple levels while maintaining the historical interaction information. The general feature enhancement task enhances the general feature from both the graph and feature perspectives to improve the robustness of our model. Extensive experiments on three publicly available datasets demonstrate the effectiveness of our method. Our code is publicly available at 
\href{https://github.com/Jinfeng-Xu/MENTOR}{https://github.com/Jinfeng-Xu/MENTOR}.

\end{abstract}

\ccsdesc[500]{Information systems~Recommender systems; Multimedia and multimodal retrieval}

\keywords{Mutimodal Recommendation, Self-supervised learning}


\maketitle

\section{Introduction}
The rapid growth of e-commerce businesses has led to significant information overload. Recommendation systems aim to alleviate information overload by simulating user preferences. However, the performance of traditional recommendation systems is limited by the data sparsity problem \cite{koren2021advances}. Recent works on multimodal recommendation mitigate this limitation by utilizing multimedia information, which can improve user preference modeling and enhance recommendation accuracy. For example, a line of work \cite{he2016vbpr,chen2017attentive} directly leverages multimodal information as side information to improve the recommendation performance. In recent years, many traditional works \cite{he2020lightgcn,wang2019neural} utilize the graph convolutional network (GCN) to capture latent information between users and items. Inspired by these works, MMGCN \cite{wei2019mmgcn} builds the user-item interaction graph for each modality separately and aggregates their prediction as the final rating prediction. DualGNN \cite{wang2021dualgnn} builds an extra homogeneous user-user graph to explore the common user preference pattern. LATTICE \cite{zhang2021mining} and FREEDOM \cite{zhou2023tale} introduce the item semantic graph to capture the latent semantically correlative signals. 

Although the GCN-based multimodal recommendation methods achieve outstanding performance, they still require sufficient high-quality historical interaction labels. Recently, a series of traditional recommendation methods leverage self-supervised learning (SSL) to alleviate label dependence. SelfCF \cite{zhou2023selfcf} uses self-supervised signals to enhance recommendation performance without relying on labels. SimGCL \cite{yu2022graph} and XSimGCL \cite{yu2023xsimgcl} effectively combine GCN and self-supervised learning to design effective self-supervised learning. SLMRec \cite{tao2022self} migrates SSL to the multimodal recommendation field, which effectively enhances the robustness of the model. 

In the multimodal recommendation field, self-supervised learning contributes to robustness enhancement while also effectively aligning features between different modalities. BM3 \cite{zhou2023bootstrap} and MMSSL \cite{wei2023multi} design self-supervised learning tasks to align modalities. However, BM3 only utilizes multimodal information as bootstrap information, the multimodal feature has still not been fully explored. MMSSL effectively aligns the features between different modalities, but results in the perturbation of historical interaction information.

To tackle the above problems, we propose a novel \underline{M}ulti-level s\underline{E}lf-supervised lear\underline{N}ing for mul\underline{T}im\underline{O}dal \underline{R}ecommendation (MENTOR). MENTOR devises a novel multilevel SSL task that improves model robustness and aligns different modality features without losing historical interaction information. In particular, we first leverage the user-item interaction graph to extract modality-specific features for each modality and then fuse the visual and textual modalities as the fused modality in the early stage. To further enhance historical interaction information, we capture semantic signals between items through item-item graphs for each modality. Furthermore, we devise the multilevel cross-modal alignment task and the general feature enhancement task. More specifically, the multilevel cross-modal alignment task uses four levels of alignment to effectively align different modalities without losing historical interaction information, and we verify the effectiveness of this task through visualization in Section~\ref{visualization}. The general feature enhancement task enhances the general features and robustness of our model on both feature masking and graph perturbation. It is expected that these general features can cover good semantic or structural meanings. In a nutshell, our contributions can be summarized as follows:
\vskip -0.1in
\begin{itemize}
    \item We propose a novel framework MENTOR for the multimodal recommendation, which alleviates both data sparsity and label sparsity problems.
    \item We propose a novel multilevel cross-modal alignment task, which effectively aligns different modality features without losing historical interaction information.
    \item We develop a general feature enhancement task, which enhances the general features on both feature masking and graph perturbation.
    \item We perform comprehensive experiments on three public datasets in Amazon to validate the effectiveness of our MENTOR on both overall and component levels.
\end{itemize}

\section{Related work}
\subsection{Multimodal Recommendation}
Many recent works incorporate multimodal information to alleviate the data sparsity problem of traditional recommendations and pursue state-of-the-art performance. VBPR \cite{he2016vbpr} is the first attempt to utilize visual content to alleviate the data sparsity problem, it considers visual content as side information to enhance the item representation based on matrix factorization \cite{rendle2012bpr}. Moreover, many works \cite{chen2019personalized,liu2019user} enhance the representation of items with both visual and textual modalities to further mitigate the data sparsity problem. To improve the fusion quality for all modalities, ACF \cite{chen2017attentive} leverages the attention mechanism to capture user preference for each modality adaptively. Inspired by the traditional recommendation system, MMGCN \cite{wei2019mmgcn} adopts GCN to construct a bipartite graph to extract the latent information in user-item interactions. GRCN \cite{wei2020graph} prunes the false-positive edges based on MMGCN to reduce the noise in the user-item bipartite graph. More recently, MKGAT \cite{sun2020multi} proposes a graph attention mechanism to balance the weights on different modalities during modal fusion. To explicitly mine the common preferences between users, DualGNN \cite{wang2021dualgnn} constructs an extra user co-occurrence graph. LATTICE \cite{zhang2021mining} introduces an item semantic graph to capture the latent correlative signals between items. FREEDOM \cite{zhou2023tale} reduces the noise from the user-item interaction graph based on LATTICE, and improves the recommendation accuracy by freezing the item semantic graph after initialization. However, existing methods still lack the ability to extract modality-specific features and modality-common features well at the same time.

\subsection{Self-supervised Learning on Recommendation}
In the traditional recommendation field, self-supervised learning (SSL) effectively improves the robustness of the model and mitigates the label dependency. SelfCF \cite{zhou2023selfcf} and BUIR \cite{lee2021bootstrapping} generate different views to learn the representation of users and items from positive interaction, respectively. To explore the negative interaction, MixGCF \cite{huang2021mixgcf} designs a general negative sampling plugin that can be directly used to train GNN-based recommender systems. SSL4Rec \cite{yao2021self} and SEPT \cite{yu2021socially} propose contrastive learning (CL) in the recommendation system field. These methods generate extra user data views via augmentation and improve each encoder with self-supervision signals from other users iteratively. MHCN \cite{yu2021self} and SGL \cite{wu2021self} propose to generate SSL signals via contrasting positive node pairs based on various augmentation operations, such as semantic neighbor identification and random walk graph sampling. SimGCL \cite{yu2022graph} proposes a simple CL method that discards the graph augmentations and instead adds uniform noises to the embedding space for creating contrastive views. More recently, XSimGCL \cite{yu2023xsimgcl} proposes a simple yet effective noise-based augmentation approach based on SimGCL, which can smoothly adjust the uniformity of the representation distribution through CL. 

In the multimodal recommendation field, SLMRec \cite{tao2022self} proposes two SSL tasks to enhance the robustness of the model, including noise perturbation over features and multimodal pattern uncovering augmentation. SSL can also be used to align features from different modalities, BM3 \cite{zhou2023bootstrap} simplifies the SSL task based on SLMRec, and proposes two tasks to align modality features under both inter-modality and intra-modality perspectives. MMSSL \cite{wei2023multi} designs a cross-modal contrastive learning task to preserve the inter-modal semantic commonality and user preference diversity. These methods inevitably generate lots of noise along with modal alignment, which leads to significant attenuation of historical interaction information. 

In this work, we propose a multilevel cross-modal alignment task, which can effectively align different modality features while retaining historical interaction information.

\begin{figure*}[h]
    \centering
    \includegraphics[width=0.95\linewidth]{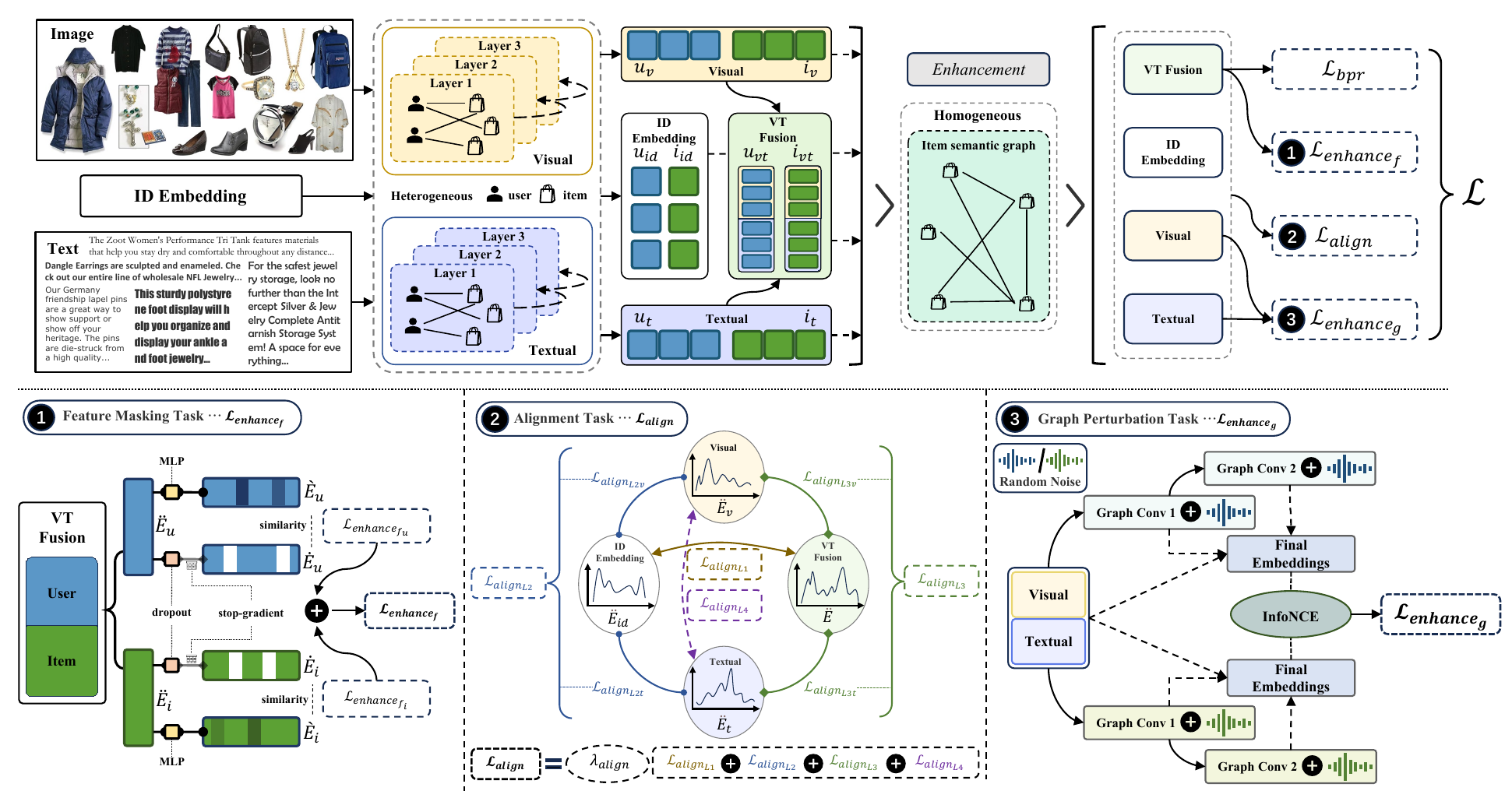}
    \vskip -0.15in
    \caption{The architecture of our MENTOR. We first utilize the graph convolutional network to extract specific features for each modality. Then, we fuse visual and textual modalities and explore the latent information with the item semantic graph based on these four modality representations of VT Fusion, ID, Visual, and Textual. We utilize an alignment self-supervised task (2) to align each modality without loss of interaction information. Besides, we leverage self-supervised tasks to enhance the general features on both the feature masking task (1) and the graph perturbation task (3).}
    \label{structure}
    \vskip -0.1in
\end{figure*}

\section{Methodology}
In this section, we present the MENTOR architecture and describe each component in our proposed model. Fig~\ref{structure} shows the overall architecture of MENTOR. 

\subsection{Problem Definition}
Let $\mathcal{U}$ = $\{u\}$ denote the user set and $\mathcal{I}$ = $\{i\}$ denote the item set. Then, we denote the features of each modality and the input ID embedding as $E_{m}$ = $\{E_{u_m} \Vert E_{i_m}\} \in \mathbb{R}^{d_m \times (|\mathcal{U}|+|\mathcal{I}|)}$, where $m \in \mathcal{M}$ is the modality, $\mathcal{M}$ is the set of modalities, $d_m$ is the dimension of the features, and $\Vert$ denotes concatenation operation. We only consider ID, visual, and textual modalities denoted by $\mathcal{M}$ = $\{id, v, t\}$. However, our model can involve more modalities than these three modalities.

\subsection{Multimodal Information Encoder}
Some previous works \cite{zhou2023tale,zhang2021mining} point out that both the user-item heterogeneous graph and the item-item homogeneous graph can significantly improve the performance of multimodal recommendations. Inspired by them, we propose a multimodal information encoder component to extract modality-specific features through the user-item graph. This component also leverages the item-item graph to capture semantically correlative signals.
\subsubsection{User-Item Graph}
To capture high-order modality-specific features, we construct three user-item graphs $\mathcal{G}$ = $\{\mathcal{G}_{m}|\mathcal{G}_{id}, \mathcal{G}_v, \mathcal{G}_t \}$. Each graph $\mathcal{G}_m$ maintains the same graph structure and only retains the node features associated with each modality. We construct all graphs inspired by LightGCN \cite{he2020lightgcn}. In particular, the user and item representations at $l$-th graph convolution layer can be formulated as:
\begin{equation}
E_{m}^{(l)}=\sum_{i \in \mathcal{N}_u} \frac{1}{\sqrt{\left|\mathcal{N}_u\right|} \sqrt{\left|\mathcal{N}_i\right|}} E_m^{(l-1)}, 
\label{func1}
\end{equation}
where $\mathcal{N}_u$ and $\mathcal{N}_i$ denote the one-hop neighbors of $u$ and $i$ in $\mathcal{G}$, respectively. 
The final embedding for each modality is calculated by element-wise summation. Formally, 
\vskip -0.05in
\begin{equation}
    \bar{E}_m = \sum_{l=0}^LE^{(l)}_m,
\end{equation}
\vskip -0.05in
where $L$ is the number of user-item graph layers.

\subsubsection{Item-Item Graph}
To extract significant semantic relations between items, we use KNN to establish the item-item graph based on the item features for each modality $m$. Particularly, we calculate the similarity score $S^m_{i,i^{\prime}}$ between item pair $(i,i^{\prime}) \in \mathcal{I}$ by the cosine similarity on their modality original features $f_i^m$ and $f_{i^{\prime}}^m$. 
\vskip -0.1in
\begin{equation}
S^m_{i,i^{\prime}}=\frac{\left(f_i^m\right)^{\top} f_{i^{\prime}}^m}{\left\|f_i^m\right\|\left\| f_{i^{\prime}}^m\right\|}.
\label{cosine}
\end{equation}
We only retain the top-$k$ neighbors with the highest similarity score:
\begin{equation}
S^m_{i,i^{\prime}} = \begin{cases}1 & \text { if } S^m_{i,i^{\prime}} \in \text { top-} k\left(S^m_{i,i^{\prime}}\right) \\ 0 & \text { otherwise }\end{cases}.
\end{equation}
Then, we aggregate multilayer neighbors on it to capture the semantic relations:
\begin{equation}
    A_{m}^{(l)} = \sum_{i^{\prime} \in \mathcal{N}_i} S^m_{i,i^{\prime}} A_{i^{\prime}_m}^{(l-1)},
\end{equation}
where $\mathcal{N}_i$ denotes the neighbors of item $i$. $A_{i^{\prime}, m}$ is the embedding of item $i^{\prime}$ in modality m. Inspired by \cite{zhou2023tale}, we freeze each item-item graph after initialization.  


\subsection{Multimodel Fusion}
To combine multiple modalities to jointly mine user preferences, we first enhance the final embedding $\bar{E}_m$ of the user-item graph based on the final embedding $A_{m}^{(l)}$ of the item-item graph for each modality. The enhanced embedding $\ddot{E}_m$ can be calculated as:
\begin{equation}
    \ddot{E}_m = \{\bar{E}_{u_m} \Vert  \bar{E}_{i_m} + A_{m}^{(l)}\},
\end{equation}
where $\Vert$ denotes the concatenation operation.

Then, we fuse visual and textual modalities with varying attention levels. 
\begin{equation}
   \ddot{E} = \{\alpha \times \ddot{E}_v \Vert  (1-\alpha) \times \ddot{E}_t\},
\end{equation}
where the attention weight $\alpha$ is a trainable parameter which we initialize to be 0.5, $\ddot{E_v}$ and $\ddot{E_t}$ are the representation of visual and textual modalities respectively, and $\ddot{E}$ is the final representation. 

It is worth noting that we discard ID embedding in the prediction score task. However, this does not mean that ID embedding is completely useless. ID embedding plays a guidance role in the self-supervised cross-modal alignment task, which is introduced in the following section.

\subsection{Multilevel Cross-Modal Alignment}
The feature distributions of different modalities are extremely different, which generates a lot of noise during the process of modality fusion. However, existing modal alignment methods \cite{wei2023multi,zhou2023bootstrap} basically perturb the historical interaction information. Therefore, we propose a multilevel cross-modal alignment component to align modalities from the perspective of data distribution using self-supervised learning. Specifically, our multilevel cross-modal alignment component has four levels, including the ID direct guidance, the ID indirect guidance, the modality direct alignment, and the modality indirect alignment levels. The ID direct guidance and ID indirect guidance levels fully utilize the historical interaction information features in the ID modality to enhance the historical interaction features in the fused modality, visual modality, and textual modality. Modality direct alignment and modality indirect alignment levels align visual and textual modalities with two different levels of self-supervised signaling.

\subsubsection{ID Direct Guidance}
In the ID direct guidance level, we align the fused modality $\ddot{E}$ and ID modality $\ddot{E}_{id}$ to enhance the importance of historical interaction. Inspired by PPMDR \cite{liu2023federated}, we adopt the Gaussian distribution to parameterize $\ddot{E}$ and $\ddot{E}_{id}$. Then, we calculate the distance between these two distributions as loss:
\begin{equation}
    \ddot{E} \sim N\left(\mu_{vt}, \sigma_{vt}^2\right), \quad
    \ddot{E}_{id} \sim N\left(\mu_{id}, \sigma_{id}^2\right),
\end{equation}
\begin{equation}
    \mathcal{L}_{align_{L1}} = |\mu_{id} - \mu_{vt}| + |\sigma_{id} - \sigma_{vt}|,
\end{equation}
where $(\mu_{id}, \sigma_{id})$ and $(\mu_{vt}, \sigma_{vt})$ characterize the Gaussian distribution for fused modality $\ddot{E}$ and ID modality $\ddot{E}_{id}$, respectively. 

\subsubsection{ID Indirect Guidance}
In the ID indirect guidance level, we align the visual modality $\ddot{E}_{v}$ and textual modality $\ddot{E}_{t}$ with ID modality $\ddot{E}_{id}$, respectively. Same as the ID direct guidance level, the final loss is calculated by:
\begin{equation}
    \ddot{E}_{v} \sim N\left(\mu_{v}, \sigma_{v}^2\right), \quad
    \ddot{E}_{t} \sim N\left(\mu_{t}, \sigma_{t}^2\right),
\end{equation}
\vskip -0.1in
\begin{equation}
    \mathcal{L}_{align_{L2v}} = |\mu_{id} - \mu_{v}| + |\sigma_{id} - \sigma_{v}|,
\end{equation}
\begin{equation}
    \mathcal{L}_{align_{L2t}} = |\mu_{id} - \mu_{t}| + |\sigma_{id} - \sigma_{t}|,
\end{equation}
\begin{equation}
    \mathcal{L}_{align_{L2}} = \mathcal{L}_{align_{L2v}} + \mathcal{L}_{align_{L2t}},
\end{equation}
where $(\mu_{v}, \sigma_{v})$ and $(\mu_{t}, \sigma_{t})$ characterize the Gaussian distribution for visual modality $\ddot{E}_{v}$ and textual modality $\ddot{E}_{t}$, respectively. 

\subsubsection{Modality Direct Alignment}
In the modality direct guidance level, we align the visual modality $\ddot{E}_{v}$ and textual modality $\ddot{E}_{t}$ with fused modality $\ddot{E}$, respectively. Different from the above two levels, this level aims to directly align the distribution of modalities to reduce modal noise in the fused modality:
\vskip -0.1in
\begin{equation}
    \mathcal{L}_{align_{L3v}} = |\mu_{vt} - \mu_{v}| + |\sigma_{vt} - \sigma_{v}|,
\end{equation}
\begin{equation}
    \mathcal{L}_{align_{L3t}} = |\mu_{vt} - \mu_{t}| + |\sigma_{vt} - \sigma_{t}|,
\end{equation}
\begin{equation}
    \mathcal{L}_{align_{L3}} = \mathcal{L}_{align_{L3v}} + \mathcal{L}_{align_{L3t}}.
\end{equation}

\subsubsection{Modality Indirect Alignment}
In the modality indirect guidance level, we align the visual modality $\ddot{E}_{v}$ and textual modality $\ddot{E}_{t}$. It aims to indirectly align the distribution of modalities, the alignment loss is defined as:
\begin{equation}
    \mathcal{L}_{align_{L4}} = |\mu_{v} - \mu_{t}| + |\sigma_{v} - \sigma_{t}|.
\end{equation}

We finally calculate the overall multilevel cross-modal alignment loss $\mathcal{L}_{align}$. Formally:
\begin{equation}
    \mathcal{L}_{align} = \lambda_{align}(\mathcal{L}_{align_{L1}} + \mathcal{L}_{align_{L2}} + \mathcal{L}_{align_{L3}} + \mathcal{L}_{align_{L4}}),
\end{equation}
where $\lambda_{align}$ is the balancing hyper-parameter.

\subsection{General Feature Enhancement}
We propose a general feature enhancement component to generate multiple views based on representation, and then capture the consistent features among these views to enhance the robustness of the recommendations. This self-supervised approach allows general features to be extracted from the raw data, thus mitigating the problem of data sparsity. Our general feature enhancement component can be divided into two tasks: feature masking and graph perturbation. The feature masking task generates different views from feature perspectives as shown in (1) of Fig~\ref{structure}. The graph perturbation task generates different views from the graph perspective using random noise as shown in (3) of Fig~\ref{structure}.

\subsubsection{Feature Masking}
We first split $\ddot{E}$ as two sides $\ddot{E}_u$ and $\ddot{E}_i$. Then, we mask out a subset of these embedding to generate contrastive views $\dot{E}_u$ and $\dot{E}_i$:
\begin{equation}
    \dot{E}_u = \ddot{E}_u \cdot \operatorname{Bernoulli(p)},
\end{equation}
\begin{equation}
    \dot{E}_i = \ddot{E}_i \cdot \operatorname{Bernoulli(p)},
\end{equation}
Inspired by BM3 \cite{zhou2023bootstrap}, we randomly mask out a subset of the representation $\ddot{E}_u$ and $\ddot{E}_i$ by dropout mechanism \cite{srivastava2014dropout}, and $p$ is the dropout ratio. 

Inspired by \cite{zhou2023bootstrap}, we place stop-gradient on the contrastive views $\dot{E}_i$ and $\dot{E}_u$. Then we transfer $\ddot{E}_u$ and $\ddot{E}_i$ through MLP.
\begin{equation}
    \grave{E}_u = \ddot{E}_u W + b,
\end{equation}
\begin{equation}
    \grave{E}_i = \ddot{E}_i W + b,
\end{equation}
where $W \in \mathbb{R}^{d\times d}$, $b \in \mathbb{R}^{d}$ denote the linear transformation matrix and bias. 

Finally, the feature masking loss is defined as:
\begin{equation}
    \mathcal{L}_{enhance_{fu}} = 1 - \operatorname{Sim}(\dot{E}_u, \grave{E}_u),
\end{equation}
\begin{equation}
    \mathcal{L}_{enhance_{fi}} = 1 - \operatorname{Sim}(\dot{E}_i, \grave{E}_i),
\end{equation}
\begin{equation}
    \mathcal{L}_{enhance_{f}} = \mathcal{L}_{enhance_{fu}} + \mathcal{L}_{enhance_{fi}},
\end{equation}
where $\operatorname{Sim}$ is the cosine similarity defined in Eq.~\ref{cosine}.
This self-supervised task aims to extract the general semantic information.

\subsubsection{Graph Perturbation}
We follow the most commonly used augmentation methods based on dropout mechanism in graphs \cite{wu2021self,you2020graph} to construct contrastive views of structural perturbations for both visual and textual modalities. We propose a perturbed user-item graph:
\begin{equation}
\dagger{E}_{m}^{(l)}=\sum_{i \in \mathcal{N}_u} \frac{1}{\sqrt{\left|\mathcal{N}_u\right|} \sqrt{\left|\mathcal{N}_i\right|}} \dagger{E}_m^{(l-1)} + \Delta^{(l)}, 
\end{equation}
where most parameters are the same as Eq.~\ref{func1}, and $\Delta^{(l)} \in \mathbb{R}^{d_m} \sim U(0,1)$ is a random noise vector.
The final perturbed embedding for each modality is calculated by element-wise summation. Formally, 
\begin{equation}
    \dagger{\bar{E}}_m = \sum_{l=0}^L\dagger{E}_{m}^{(l)}.
\end{equation}
For both visual and textual modalities, we generate two contrastive views $\dagger{\bar{E}}_m^1$ and $\dagger{\bar{E}}_m^2$ for each modality and adopt InfoNCE \cite{oord2018representation} for contrastive learning. Formally, the graph perturbation loss for each modality is defined as:
\begin{equation}
\begin{aligned}
\mathcal{L}_{enhance_{g_m}} & =\sum_{u \in \mathcal{U}}-\log \frac{\exp \Big(e_{u,m}^1 \cdot e_{u,m}^2 / \tau\Big)}{\sum_{v \in \mathcal{U}} \exp \Big(e_{v,m}^1 \cdot e_{v,m}^2 / \tau\Big)} \\
& +\sum_{i \in I}-\log \frac{\exp \Big(e_{i,m}^1 \cdot e_{i,m}^2 / \tau\Big)}{\sum_{j \in I} \exp \Big(e_{j,m}^1 \cdot e_{j,m}^2 / \tau\Big)},
\end{aligned}
\end{equation}
where $e_{u/v,m}^1$ and $e_{u/v,m}^2$ are the modality $m$ features of user $u/v$ in contrastive views $\dagger{\bar{E}}_m^1$ and $\dagger{\bar{E}}_m^2$. Besides, $e_{i/j,m}^1$ and $e_{i/j,m}^2$ are the modality $m$ features of item $i/j$ in contrastive views $\dagger{\bar{E}}_m^1$ and $\dagger{\bar{E}}_m^2$. $\tau$ is the temperature hyper-parameter of softmax.

The total graph perturbation loss is calculated as:
\begin{equation}
    \mathcal{L}_{enhance_{g}} = \mathcal{L}_{enhance_{g_v}} + \mathcal{L}_{enhance_{g_t}}.
\end{equation}
This self-supervised task aims to extract the general structural meaning of user-item interaction.

Finally, the overall general feature enhancement loss is:
\begin{equation}
    \mathcal{L}_{enhance} = \lambda_{g}\mathcal{L}_{enhance_{g}} + \lambda_{f}\mathcal{L}_{enhance_{f}},
\end{equation}
where $\lambda_{g}$ and $\lambda_{f}$ are the balancing hyper-parameters.

\subsection{Optimization}
We adopt the Bayesian Personalized Ranking (BPR) loss \cite{rendle2012bpr} as the basic optimization function.  Essentially, BPR aims to widen the predicted preference margin between the positive and negative items for each triplet $(u, p, n) \in \mathcal{D}$, where $\mathcal{D}$ denotes the training set. The positive item $p$ refers to the one with which the user $u$ has interacted, while the negative item $n$ has been randomly chosen from the set of items that the user $u$ has not interacted with. The BPR function is defined as follow:
\begin{equation}
    \mathcal{L}_{bpr} = \sum_{(u, p, n) \in \mathcal{D}} - \log(\sigma(y_{u,p} - y_{u,n})),
\end{equation}
where $y_{u,p}$ and $y_{u,n}$ are the ratings of user $u$ to the positive item $p$ and negative item $n$, calculated by $\ddot{E}_u^\mathrm{T}\ddot{E}_p$ and $\ddot{E}_u^\mathrm{T}\ddot{E}_n$, respectively. $\sigma$ is the Sigmoid function.

Then we update the representation of users and items by the combination of BPR loss, multilevel cross-modal alignment loss, and general feature enhancement loss. Formally:
\begin{equation}
    \mathcal{L} = \mathcal{L}_{bpr} + \mathcal{L}_{align} + \mathcal{L}_{enhance} + \lambda_{E}(\|E_v\|_2^2 + \|E_t\|_2^2),
\end{equation}
where $E_v$ and $E_t$ are the model parameters. $\lambda$ is a hyperparameter to control the effect of the $L_2$ regularization.

\begin{table}[!ht]
\vskip -0.1in
    \centering
\caption{Statistics of the experimental datasets}
\vskip -0.05in
\label{tab:dataset_statistics}
    \begin{tabular}{ccccc}
    \hline
         Dataset&  \# Users&  \# Items&  \# Interaction& Sparsity\\
         \hline
         Baby &  19445 &  7050 &  160792 & 99.88\%\\
         Sports &  35598 &  18357 &  296337 & 99.95\%\\
         Clothing &  39387 &  23033 &  278677 & 99.97\%\\
         \hline
    \end{tabular}
\vskip -0.15in
\end{table}

\section{Experiment}
In this section, we conduct comprehensive experiments to evaluate the performance of our MENTOR model on three widely used real-world datasets. The following four questions can be well answered through experiment results:
\begin{itemize}
    \item RQ1: How effective is our MENTOR compared with the state-of-the-art traditional recommendation methods and multimedia recommendation methods?
    \item RQ2: How do the key components impact the performance of our MENTOR?
    \item RQ3: Can the multilevel cross-modal alignment component effectively align different modalities?
    \item RQ4: How do different hyper-parameter settings influence the performance of our MENTOR?
\end{itemize}
\subsection{Experimental Settings}

\subsubsection{Datasets}
To evaluate our proposed MENTOR in the top-N item recommendation task, we conduct extensive experiments on three categories of the widely used Amazon dataset \cite{mcauley2015image}: (1) Baby, (2) Sports and Outdoors (denoted by Sports), and (3) Clothing, Shoes, and Jewelry (denoted by Clothing). These datasets provide both product descriptions and images simultaneously. Following the previous works \cite{he2016vbpr,wei2019mmgcn}, the raw data of each dataset are pre-processed with a 5-core setting on both items and users. Besides, we use the pre-extracted 4096-dimensional visual features and extract 384-dimensional textual features using a pre-trained sentence transformer \cite{zhou2023mmrecsm}. The statistics of these datasets are presented in Table~\ref{tab:dataset_statistics}.

\subsubsection{Baselines}
To demonstrate the effectiveness of our proposed MENTOR, we compare it with the following state-of-the-art recommendation methods, which can be divided into two groups: traditional recommendation methods and multimedia recommendation methods. 

\noindent1) Traditional recommendation methods:

\textbf{MF-BPR} \cite{rendle2012bpr}: This method leverages Bayesian personalized ranking (BPR) loss to improve the performance based on matrix factorization.

\textbf{LightGCN} \cite{he2020lightgcn}: This method simplifies the unnecessary components based on graph convolutional networks (GCN).

\textbf{LayerGCN} \cite{zhou2023layer}: This method proposes a layer-refined GCN, which utilizes residual connection to alleviate the over-smoothing problem for the GCN-based methods.

 \noindent2) Multimedia recommendation methods:

\textbf{VBPR} \cite{he2016vbpr}: This method utilizes visual features as side information to enhance the performance of MF-BPR. To ensure the fairness of our evaluation, we extend VBPR with textual modality.

\textbf{MMGCN} \cite{wei2019mmgcn}: This method constructs a specific graph for each modality utilizing GCN. The final prediction results are calculated by fusing the results for all modalities.

\textbf{DualGNN} \cite{wang2021dualgnn}: This method constructs an extra user co-occurrence graph to explore the common preference pattern.

\textbf{LATTICE} \cite{zhang2021mining}: This method constructs an extra item semantic graph to capture the latent semantically correlative signals.

\textbf{FREEDOM} \cite{zhou2023tale}: This method denoises the user-item graph based on LATTICE, and freezes the item-item graph to improve recommendation performance.

\textbf{SLMRec} \cite{tao2022self}: This method first utilizes self-supervised learning (SSL) for the multimodal recommendation, which proposes two tasks, including noise perturbation over features and multimodal pattern uncovering augmentation.

\textbf{BM3} \cite{zhou2023bootstrap}: This method simplifies the SSL task for multimodal recommendation. It perturbs the representation through a dropout mechanism directly.

\textbf{MMSSL} \cite{wei2023multi}: This method designs a modality-aware interactive structure learning paradigm via adversarial perturbations, and proposes a cross-modal comparative learning method to disentangle the common and specific features among modalities.

\subsubsection{Evaluation Protocols}
To evaluate the performance fairly, we adopt two widely used metrics: Recall@$K$ (R@$K$) and NDCG@$K$ (N@$K$). We report the average metrics of all users in the test dataset under both $K$ =  10 and $K$ = 20. We follow the popular evaluation setting \cite{zhou2023tale} with a random data splitting 8:1:1 for training, validation, and testing.

\subsubsection{Implementation Details}
We implement our proposed MENTOR and all the baselines with MMRec \cite{zhou2023mmrecsm}. For the general settings, we initialized the embedding with Xavier initialization \cite{glorot2010understanding} of dimension 64. Besides, we optimize all models with Adam optimizer \cite{kingma2014adam}. To achieve a fair evaluation, we perform a complete grid search for each baseline method following its published paper to find the optimal setting. For our proposed MENTOR, we perform a grid search on the dropout ratio in \{0.1, 0.2, 0.3, 0.4, 0.5, 0.6, 0.7\}, balancing hyper-parameter $\lambda_f$ in \{0.5, 1, 1.5, 2, 2.5\}, balancing hyper-parameter $\lambda_g$ in \{1e-2, 1e-3, 1e-4\}, temperature hyper-parameter $\tau$ in \{0.1, 0.2, 0.3, 0.4, 0.5, 0.6, 0.7, 0.8\}, and balancing hyper-parameter $\lambda_{align}$ in \{0.1, 0.2, 0.3\}. We fix the learning rate with $1e-4$, and the number of GCN layers in the heterogenous graph with $L$ = 2. The $k$ of top-$k$ in the item-item graph is set as 40. For convergence consideration, the early stopping is fixed at 20. Following \cite{zhou2023mmrecsm}, we update the best record by utilizing Recall@20 on the validation dataset as the indicator.

\begin{table*}[!ht]
    \centering
\caption{Performance comparison of Baselines and MENTOR in terms of Recall@K (R@K), and NDCG@K (N@K).}
    \vskip -0.1in
\label{tab:comparison results}
    \begin{tabular}{ccccc|cccc|cccc}
    \hline
         $^1$Datasets&  \multicolumn{4}{c}{Baby}&  \multicolumn{4}{c}{Sports}&  \multicolumn{4}{c}{Clothing}\\\hline\hline
         Model Source& R@10& R@20& N@10& N@20& R@10& R@20& N@10& N@20& R@10& R@20& N@10& N@20\\\hline
         MF-BPR & 0.0357& 0.0575& 0.0192& 0.0249& 0.0432& 0.0653& 0.0241& 0.0298 & 0.0187 & 0.0279 & 0.0103 & 0.0126\\
         LightGCN & 0.0479& 0.0754& 0.0257& 0.0328& 0.0569& 0.0864& 0.0311& 0.0387 & 0.0340 & 0.0526 & 0.0188 & 0.0236\\
         LayerGCN & \underline{0.0529}& \underline{0.0820}& \underline{0.0281}& \underline{0.0355}& \underline{0.0594}& \underline{0.0916}& \underline{0.0323}& \underline{0.0406}& \underline{0.0371}& \underline{0.0566}& \underline{0.0200}& \underline{0.0247}\\
         MENTOR &  \textbf{0.0678}&  \textbf{0.1048}& \textbf{0.0362}&  \textbf{0.0450}&  \textbf{0.0763}&  \textbf{0.1139} &\textbf{0.0409} & \textbf{0.0511} 
         &  \textbf{0.0668}&  \textbf{0.0989} &\textbf{0.0360} & \textbf{0.0441} \\\hline
         Improv.&  28.17\%&  27.80\%&  28.83\%&  26.76\%&  28.45\%&  24.34\%& 26.63\%&  25.86\% &  80.05\%&  74.73\%&  80.00\%&  78.54\%\\\hline\hline
         VBPR &  0.0423&  0.0663&  0.0223&  0.0284&  0.0558&   0.0856& 0.0307& 0.0384& 0.0281& 0.0415& 0.0158& 0.0192\\
         MMGCN &  0.0378&  0.0615&  0.0200&  0.0261&  0.0370&   0.0605& 0.0193& 0.0254 & 0.0218& 0.0345& 0.0110& 0.0142\\
         DualGNN &  0.0448&  0.0716&  0.0240&  0.0309&  0.0568&   0.0859& 0.0310& 0.0385& 0.0454& 0.0683& 0.0241& 0.0299\\
         LATTICE &  0.0547&  0.0850&  0.0292&  0.0370&  0.0620&  0.0953&  0.0335&  0.0421& 0.0492& 0.0733& 0.0268& 0.0330\\
         FREEDOM &\underline{0.0627}&\underline{0.0992}&\underline{0.0330}&\underline{0.0424}&\underline{0.0717}&\underline{0.1089}&\underline{0.0385}&\underline{0.0481}&\underline{0.0629}&\underline{0.0941}&\underline{0.0341}&\underline{0.0420}\\
         SLMRec & 0.0529& 0.0775& 0.0290& 0.0353& 0.0663&   0.0990& 0.0365& 0.0450 &0.0452& 0.0675& 0.0247& 0.0303\\
         BM3 & 0.0564& 0.0883& 0.0301& 0.0383& 0.0656& 0.0980& 0.0355& 0.0438& 0.0422& 0.0621& 0.0231& 0.0281\\
         MMSSL & 0.0613& 0.0971& 0.0326& 0.0420& 0.0673& 0.1013& 0.0380& 0.0474& 0.0531& 0.0797& 0.0291& 0.0359\\
         MENTOR &  \textbf{0.0678}&  \textbf{0.1048}& \textbf{0.0362}&  \textbf{0.0450}&  \textbf{0.0763}&  \textbf{0.1139} &\textbf{0.0409} & \textbf{0.0511} 
         &  \textbf{0.0668}&  \textbf{0.0989} &\textbf{0.0360} & \textbf{0.0441} 
         \\\hline
         Improv.&  8.13\%&  5.64\%&  9.69\%&  6.13\%&  6.42\%&   4.59\%& 6.23\%& 6.24\% &  6.20\%&   5.10\%& 5.57\%& 5.00\% \\\hline\hline
    \end{tabular}
    \\ $^1$Datasets can be accessed at \href{http://jmcauley.ucsd.edu/data/amazon/links.html}{http://jmcauley.ucsd.edu/data/amazon/links.html}.
\end{table*}

\subsection{Effectiveness of MENTOR (RQ1)}
Table~\ref{tab:comparison results} summarizes the performance of our MENTOR and other baselines on three datasets. We find the following observations:

\textbf{Our MENTOR achieves better performance than both traditional recommendation methods and multimodal recommendation methods.} More specifically, we first utilize a multimodal information encoder to enrich modality features by encoding the high-order relations between users and items and semantically correlative signals between items. Besides, previous self-supervised methods merely utilize multimodal information such as textual and visual to play a bootstrap role. They still use traditional ID embedding for the final prediction. Instead, our work just utilizes ID embedding to guide feature alignment between modalities, and we directly use textual and visual modalities to participate in the final prediction. This effectively reduces noise from the modality fusion process. We then enhance the general features through two self-supervised learning, including the feature masking task and the graph perturbation task. This improves the robustness of our method. As a result, our MENTOR improves the best baseline in the traditional recommendation methods group by 27.80\%, 24.34\%, and 74.73\% in terms of Recall@20 on Baby, Sports, and Clothing, respectively. Besides, our MENTOR improves the best baseline in the multimodal recommendation methods group by 5.64\%, 24.34\%, and 74.73\% in terms of Recall@20 on Baby, Sports, and Clothing, respectively. The results verify the effectiveness of our MENTOR.
 
\textbf{Leveraging multimodal information can effectively improve recommendation accuracy.} Specifically, the recent multimodal recommendation methods generally outperform traditional recommendation methods in all scenarios. This demonstrates that multimodal recommendation methods can effectively mitigate the data sparsity problem by leveraging multiple modalities to jointly mine user preferences. It is worth noting that this trend is particularly significant on the clothing dataset, where our work improves recommendation performance by around 80\% on the clothing dataset compared to LayerGCN. It suggests that visual and textual information play an extremely significant role in user preference on clothing.

\textbf{Combining homogeneous graph with the self-supervised task can significantly improve the recommendation performance}. Most previous self-supervised multimodal recommendation methods only utilize the user-item heterogeneous graph. We observe that some works such as FREEDOM, and DualGNN achieve outstanding results by adding a homogenous graph to capture the co-occurrence signals between users or between items. Our MENTOR significantly improves the accuracy for recommendation by combining self-supervised learning with homogenous graph masterly. 

\subsection{Ablation Study (RQ2)}
In this section, we conduct exhaustive experiments to evaluate the effectiveness of different components of MENTOR. 

\subsubsection{Effect of multilevel cross-modal alignment}
\label{sec:Effect of multilevel cross-modal alignment}
\begin{figure}
\vskip -0.2in
    \centering
    \includegraphics[width=0.99\linewidth]{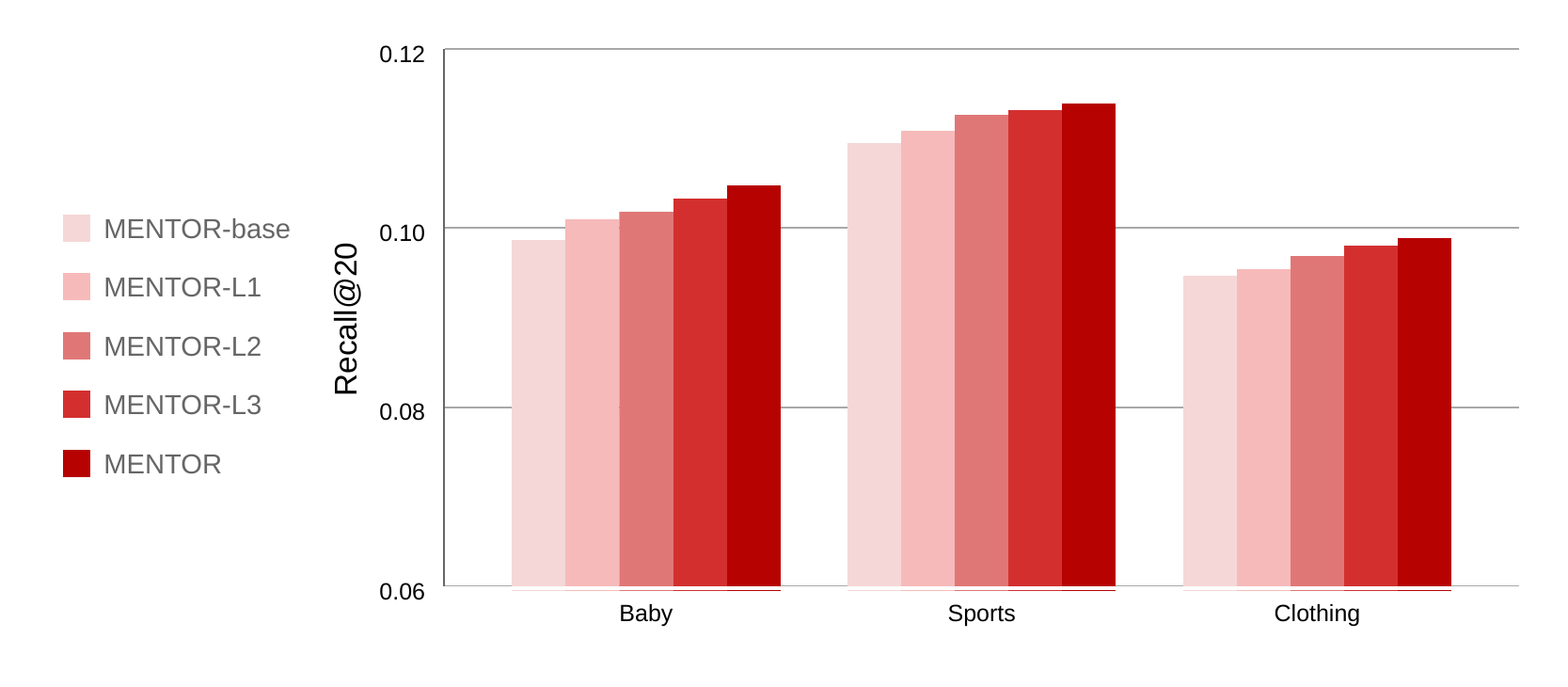}
    \vskip -0.2in
    \caption{Effect of multilevel cross-modal alignment.}
    \label{fig:multilevel cross-modal alignment}
\vskip -0.2in
\end{figure}

To investigate the effects of the multilevel cross-modal alignment component, we design the following variants of MENTOR.
\begin{itemize}[leftmargin=*]
    \item \textbf{MENTOR$_{base}$}: In this variant, we remove the multilevel cross-modal alignment component from MENTOR.
    \item \textbf{MENTOR$_{L1}$}: In this variant, MENTOR only retains the ID direct guidance part from the multilevel cross-modal alignment component.
    \item \textbf{MENTOR$_{L2}$}: In this variant, MENTOR retains the ID direct guidance and ID indirect guidance parts from the multilevel cross-modal alignment component.
    \item \textbf{MENTOR$_{L3}$}: In this variant, MENTOR retains the ID direct guidance, ID indirect guidance, and modality direct alignment parts from the multilevel cross-modal alignment component.
\end{itemize}
As shown in Fig.~\ref{fig:multilevel cross-modal alignment}, the result MENTOR$_{base}$ < MENTOR$_{L1}$ < MENTOR$_{L2}$ < MENTOR$_{L3}$ < MENTOR shows that each level of our multilevel cross-modal alignment component leads to an obvious improvement in recommendation performance, and their effects can be superimposed on each other. We will further demonstrate the effectiveness of our multilevel cross-modal alignment component by visualization in Section~\ref{visualization}.

\subsubsection{Effect of general feature enhancement}

\begin{table}
    \centering
\caption{Fine-grained performance comparison of each part of general feature enhancement component.}
\vskip -0.1in
\label{tab:general feature enhancement}
    \begin{tabular}{ll|ccc}\hline
        \multirow{2}{*}{\textbf{Variants}}& \multirow{2}{*}{\textbf{Metrics}}& \multicolumn{3}{c}{\textbf{Datasets}}\\
        &&Baby& Sports&Clothing\\ \hline\hline
        \multirow{2}{*}{MENTOR$_{fg}$}&R@20 &0.1011& 0.1094&0.0949\\
        & N@20& 0.0440& 0.0481&0.0418\\ \hline
        \multirow{2}{*}{MENTOR$_{f}$} &R@20&0.1037& 0.1126&0.0981\\ 
        &N@20&0.0449& 0.0505& \textbf{0.0443}\\\hline
        \multirow{2}{*}{MENTOR$_{g}$} & R@20 & 0.1034& 0.1129& 0.0977\\
        & N@20 & \textbf{0.0454}& 0.0507& 0.0438\\\hline
        \multirow{2}{*}{MENTOR} & R@20 & \textbf{0.1048}& \textbf{0.1139}& \textbf{0.0989}\\ 
        & N@20& 0.0450& \textbf{0.0511}& 0.0441\\\hline
    \end{tabular}
\vskip -0.15in
\end{table}
We conduct this experiment to justify the importance of the general feature enhancement component in our MENTOR with details. To verify the validity of each part of the general feature enhancement component at a fine-grained level, the following variants were constructed:
\vskip -0.1in
\begin{itemize}[leftmargin=*]
    \item \textbf{MENTOR$_{fg}$}: We remove the whole general feature enhancement component.
    \item \textbf{MENTOR$_f$}: We remove the feature masking task in the general feature enhancement component.
    \item \textbf{MENTOR$_g$}: We remove the graph perturbation task in the general feature enhancement component.
\end{itemize}
As illustrated in Table~\ref{tab:general feature enhancement}, we find that both feature masking task and graph perturbation task can effectively improve the recommendation accuracy in all datasets. Besides, these two tasks can improve recommendation accuracy together to achieve higher performance than each single task.

\begin{figure}[h]
\vskip -0.2in
    \centering
    \subfigure[distribution on MENTOR-base] {
        \label{fig:2D-1}
        \includegraphics[width=0.45\linewidth]{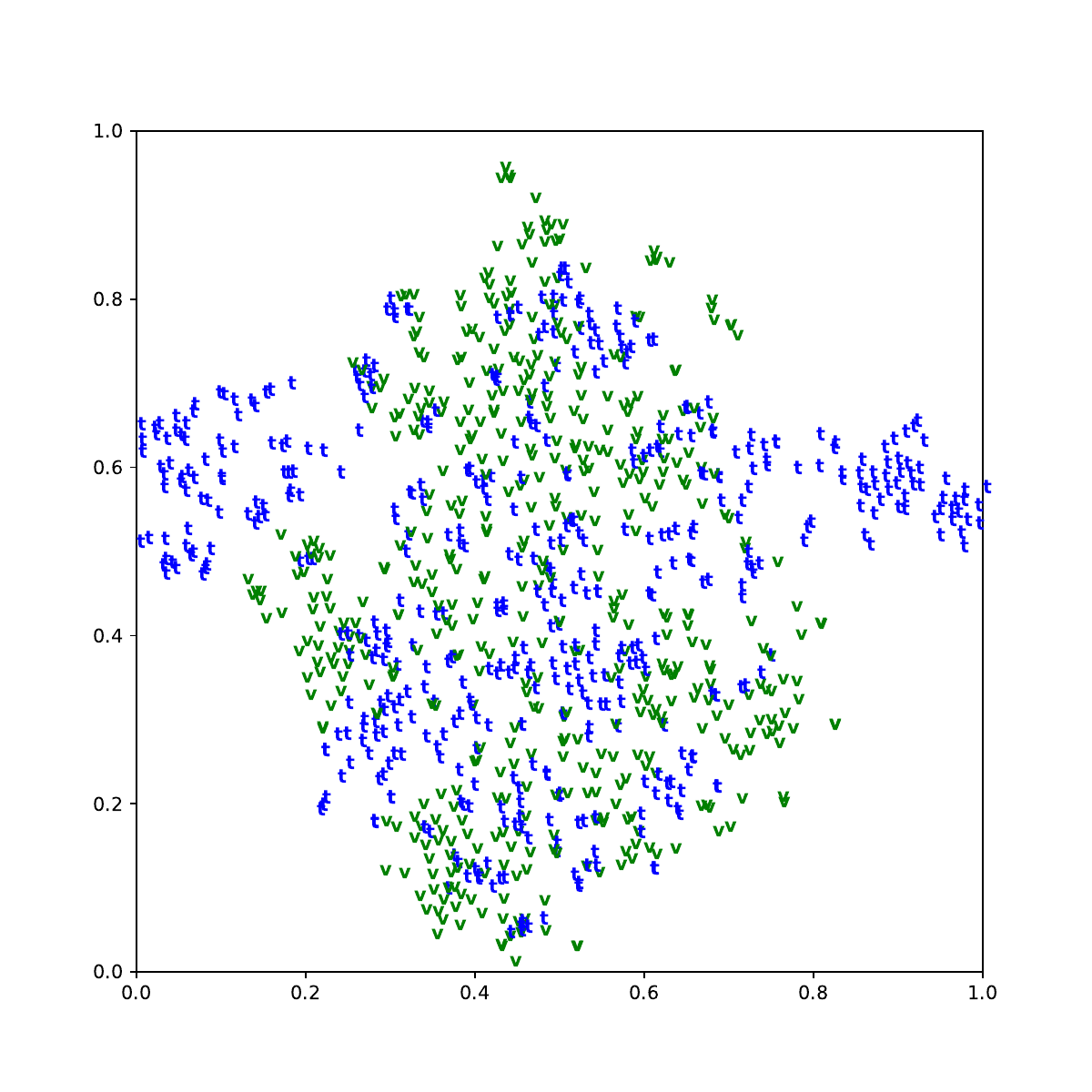}
        }   \hspace{-5mm}
    \subfigure[distribution on MENTOR] {
        \label{fig:2D-2}
        \includegraphics[width=0.45\linewidth]{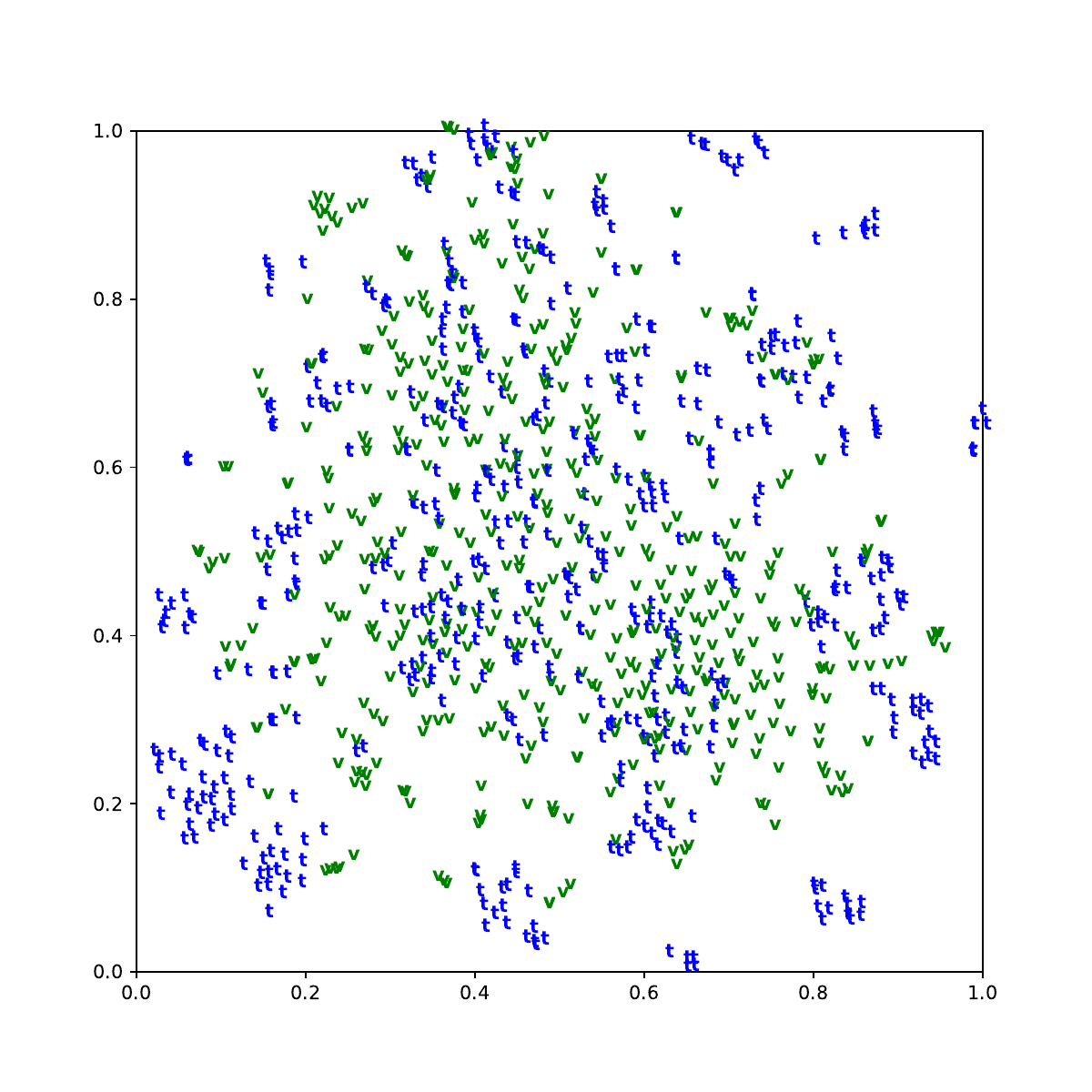}
        } 
            \vskip -0.15in
    \subfigure[density on MENTOR-base] {
        \label{fig:1D-1}
        \includegraphics[width=0.45\linewidth]{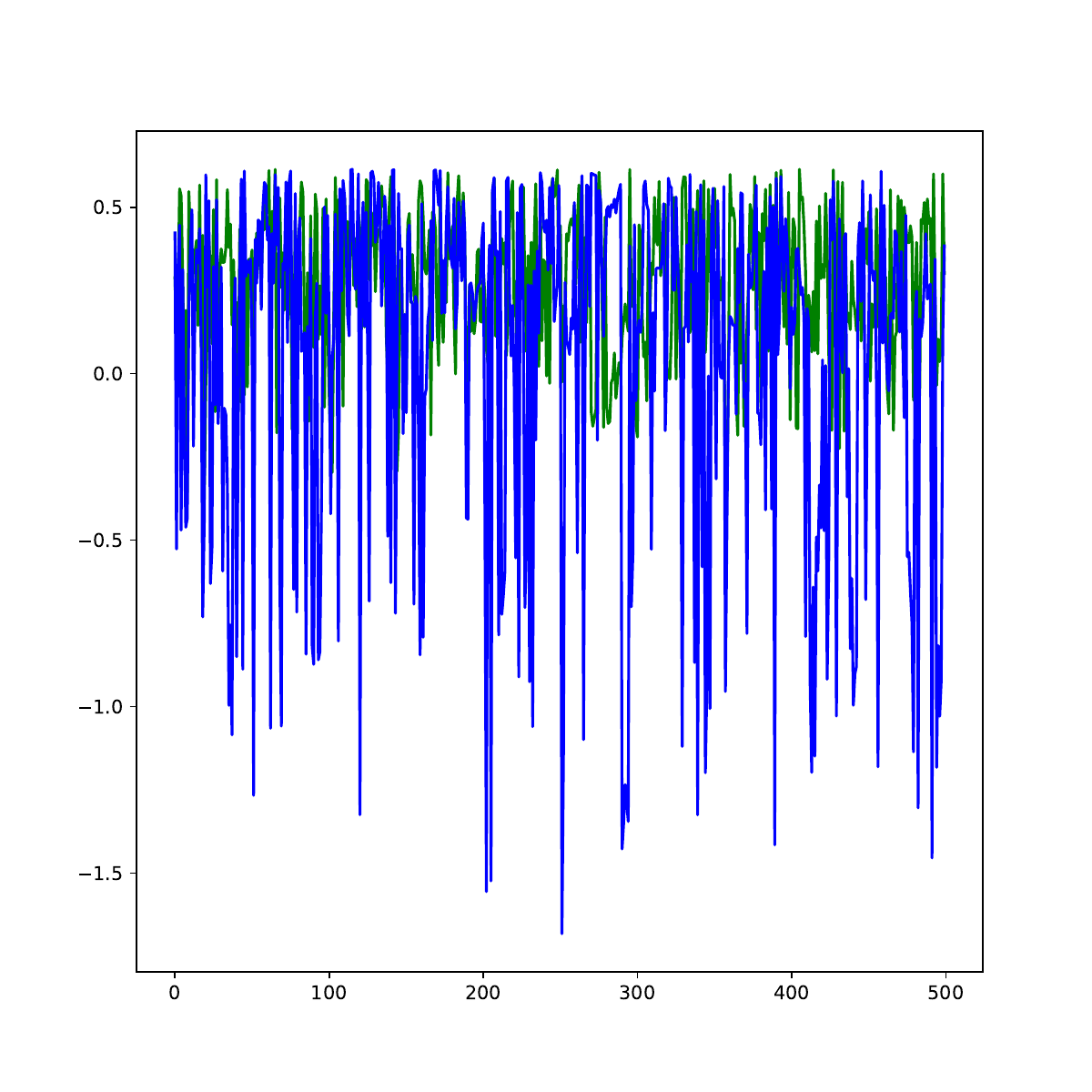}
        }   \hspace{-5mm}
    \subfigure[density on MENTOR] {
        \label{fig:1D-2}
        \includegraphics[width=0.45\linewidth]{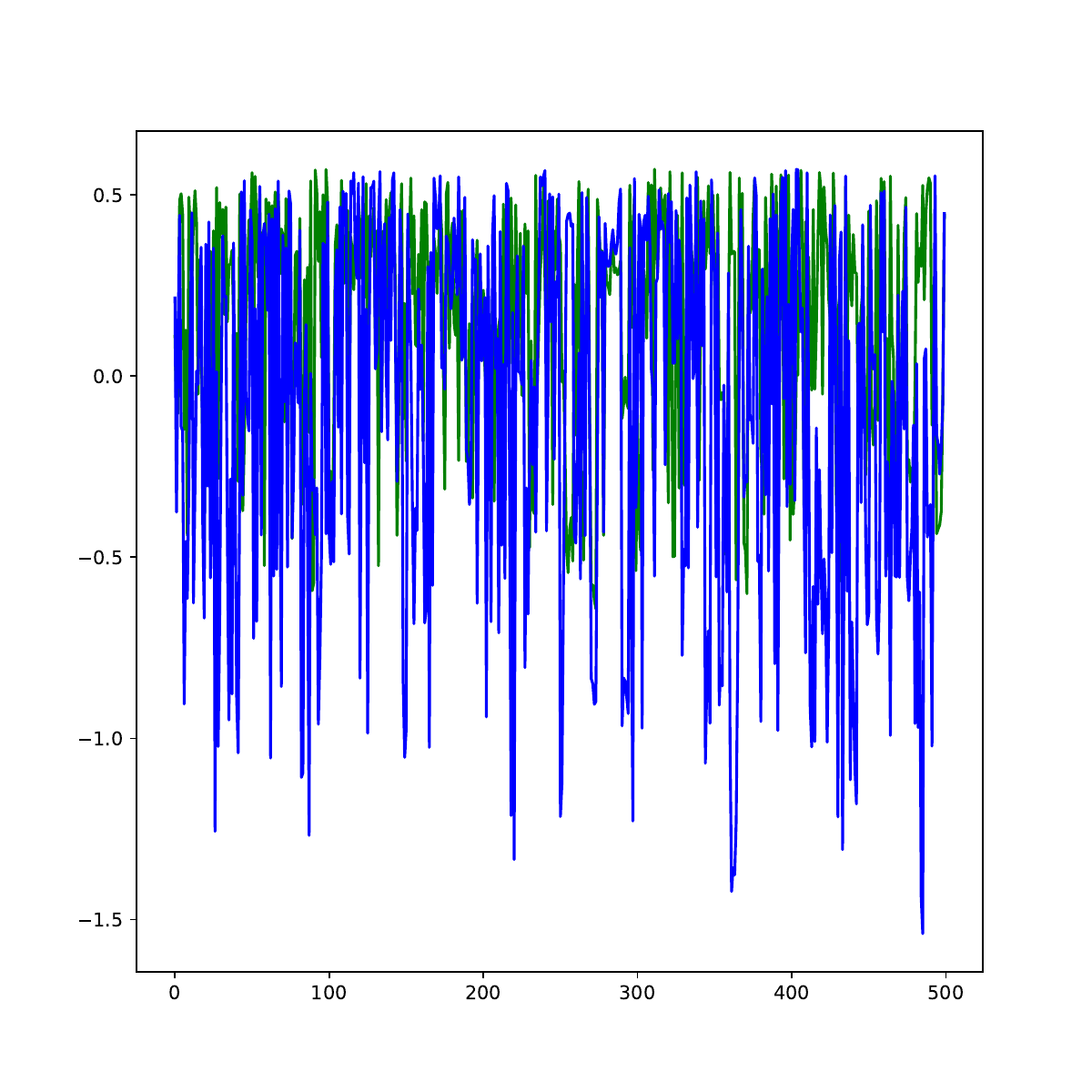}
        } 
    \vskip -0.2in
    \caption{The distribution of representations includes textual and visual modalities. Figure (a) and (c) show the distribution of MENTOR$_{base}$, and Figure (b) and (d) show the distribution of MENTOR. Blue represents the textual modality and green represents the visual modality.}   
    \label{Metrics based on normal}   
\vskip -0.2in
\end{figure}

\subsection{Visualization Analysis (RQ3)}
\label{visualization}
To further verify the effectiveness of our multilevel cross-modal alignment component, we visualize the distribution of representation in the Baby dataset. Fig.~\ref{fig:2D-1}-\ref{fig:2D-2} illustrates the impact of our multilevel cross-modal alignment component in the 2-dimension perspective. The two models of our comparison are MENTOR$_{base}$ and MENTOR, which we mentioned in Section~\ref{sec:Effect of multilevel cross-modal alignment}. To make a fair comparison, we choose the visual and textual representations of them with optimal hyper-parameters. Specifically, we randomly sample 500 items from the Baby dataset. Then, we utilize t-SNE \cite{van2008visualizing} to map their embedding to the 2-dimension space. We can observe that the textual modality distribution of MENTOR$_{base}$ is relatively more discrete than the visual modality distribution of MENTOR$_{base}$. Besides, the textual modality distribution and the visual modality distribution of MENTOR are much closer to each other compared with MENTOR$_{base}$.

To present our results more clearly, we leverage kernel density estimation (KDE) \cite{terrell1992variable} to plot the 2D feature distribution. We use a Gaussian kernel function with a bandwidth of 0.2. Fig.~\ref{fig:1D-1}-\ref{fig:1D-2} shows that our multilevel cross-modal alignment component can effectively align visual modality and textual modality. It is worth noting that we find textual modality to be more discrete than visual modality. This phenomenon is consistent with the human intuition that text encompasses a wider range of information than visual information.

\begin{figure}[h]
\vskip -0.2in
    \centering
    \subfigure[Recall@20] {
        \label{fig:recall}
        \includegraphics[width=0.40\linewidth]{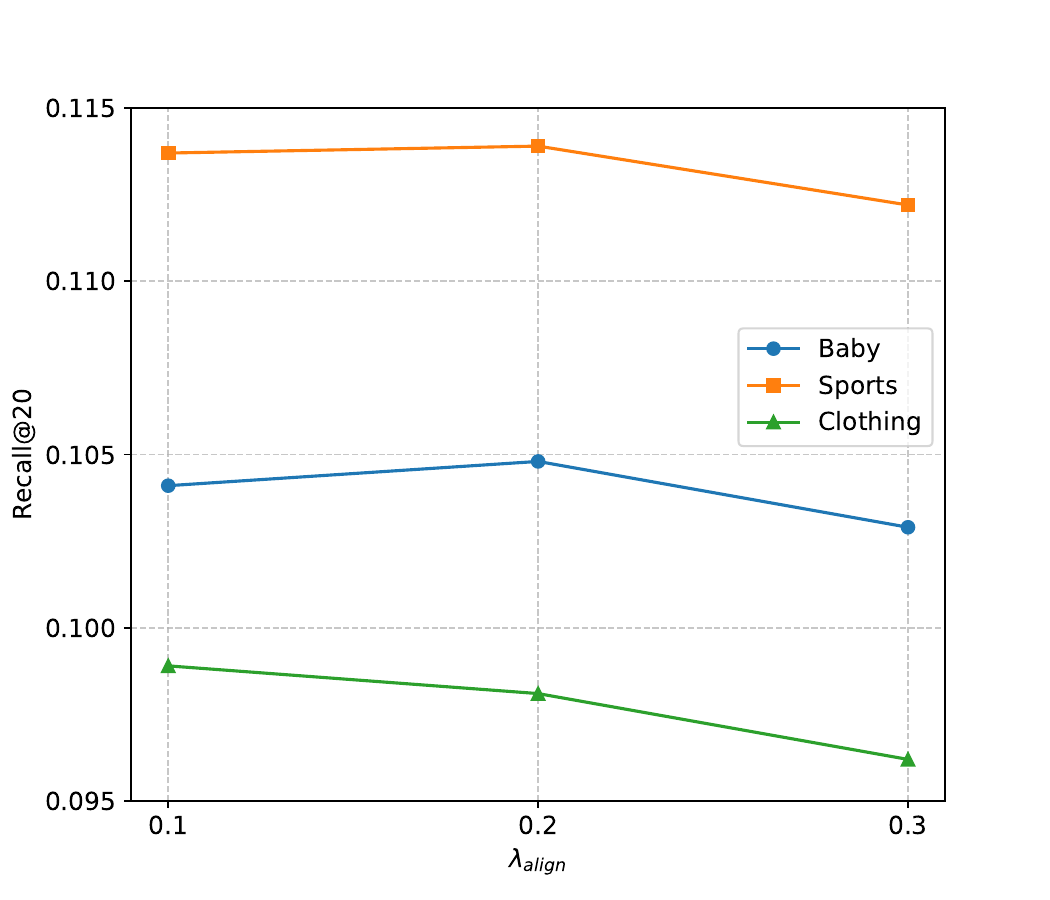}
        }
    \subfigure[NDCG@20] {
        \label{fig:ndcg}
        \includegraphics[width=0.40\linewidth]{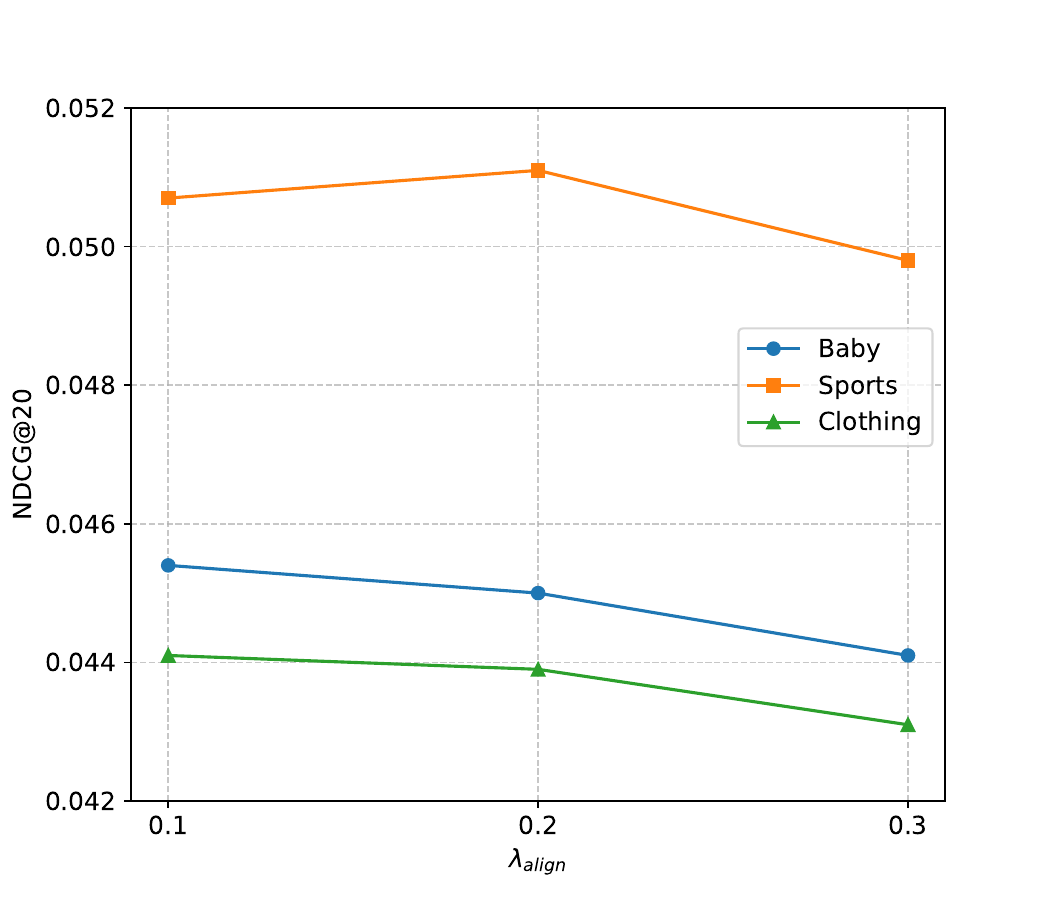}
        }
        \vskip -0.15in
    \caption{Effect of the balancing hyper-parameter $\lambda_{align}$.}
    \label{}  
\vskip -0.2in
\end{figure}



\begin{figure}[h]
\vskip -0.1in
    \centering
    \subfigure[$p$ and $\lambda_f$ on Baby] {
        \label{fig:heatmapBaby}
        \includegraphics[width=0.3\linewidth]{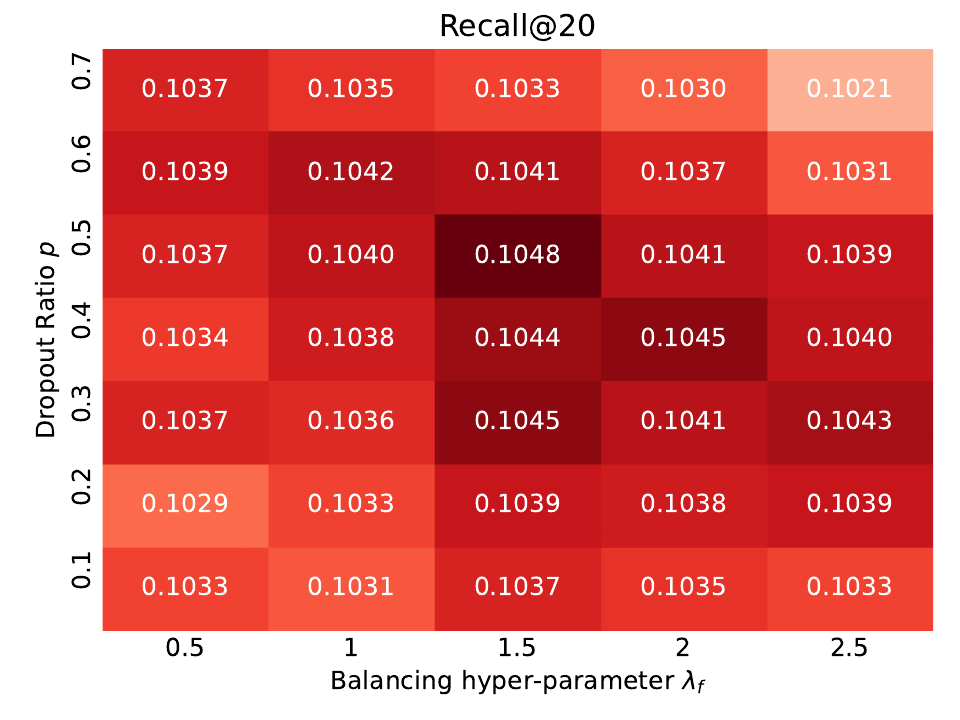}
        }
    \subfigure[$p$ and $\lambda_f$ on Sports] {
        \label{fig:heatmapSports}
        \includegraphics[width=0.3\linewidth]{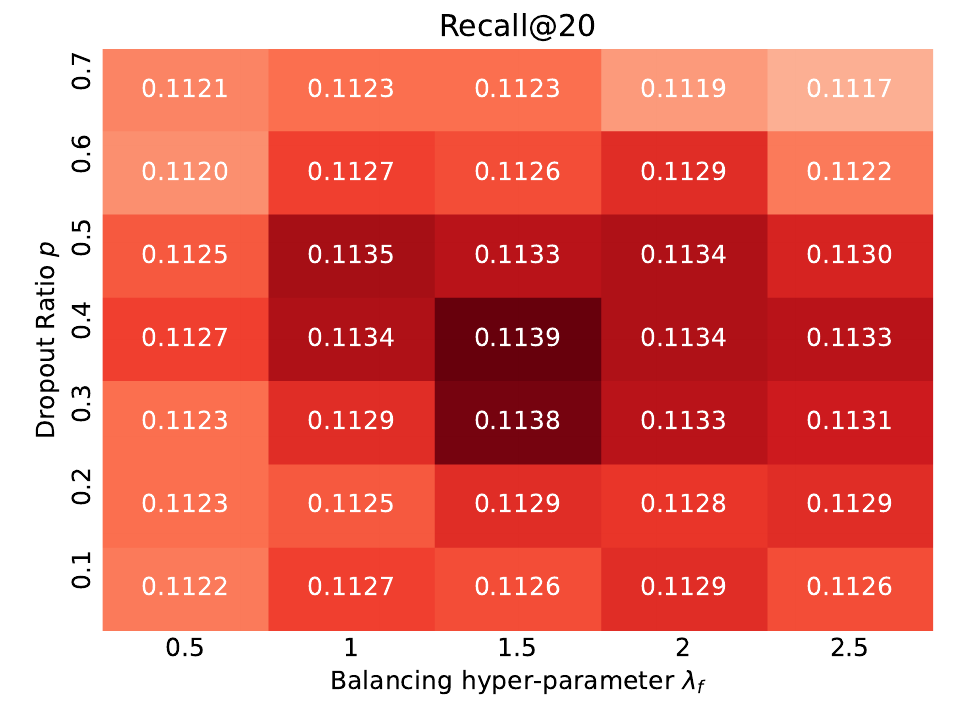}
        }  
    \subfigure[$p$ and $\lambda_f$ on Clothing] {
        \label{fig:heatmapClothing}
        \includegraphics[width=0.3\linewidth]{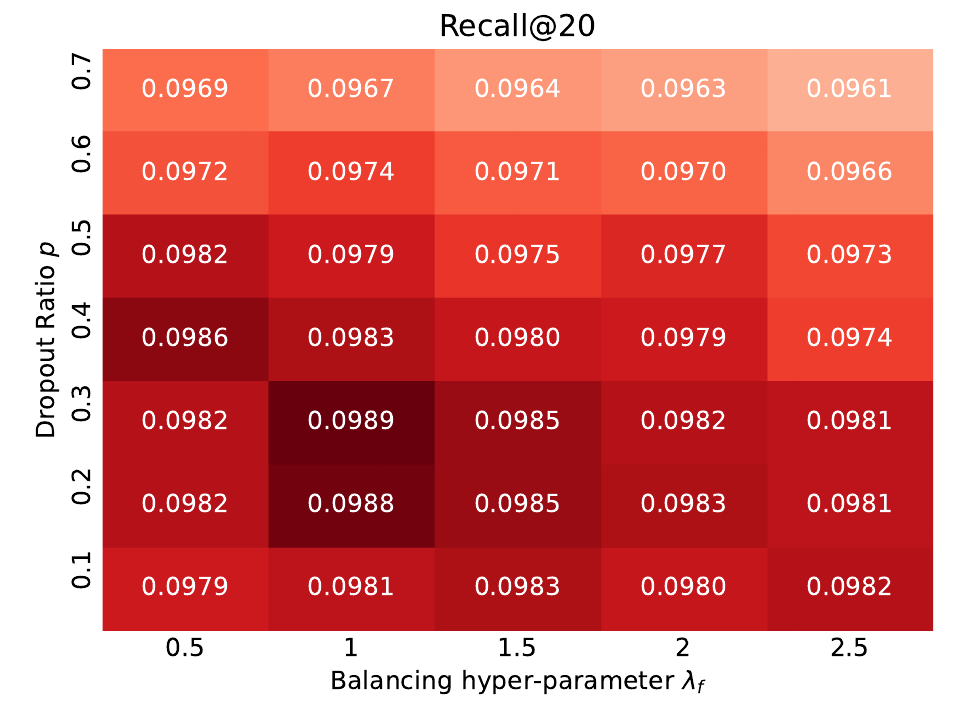}
        }    \vskip -0.1in
    \subfigure[$\tau$ and $\lambda_g$ on Baby] {
        \label{fig:heatmapBaby2}
        \includegraphics[width=0.3\linewidth]{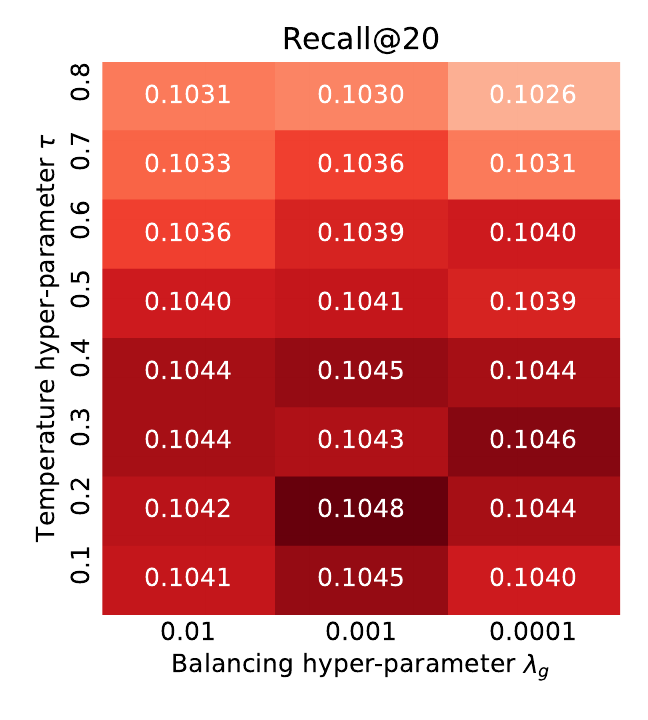}
        }
    \subfigure[$\tau$ and $\lambda_g$ on Sports] {
        \label{fig:heatmapSports2}
        \includegraphics[width=0.3\linewidth]{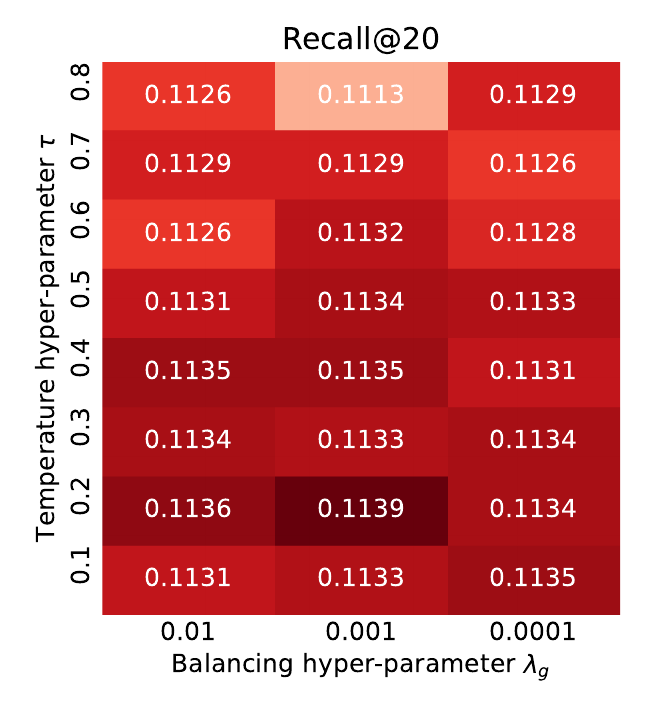}
        }
    \subfigure[$\tau$ and $\lambda_g$ on Clothing] {
        \label{fig:heatmapClothing2}
        \includegraphics[width=0.3\linewidth]{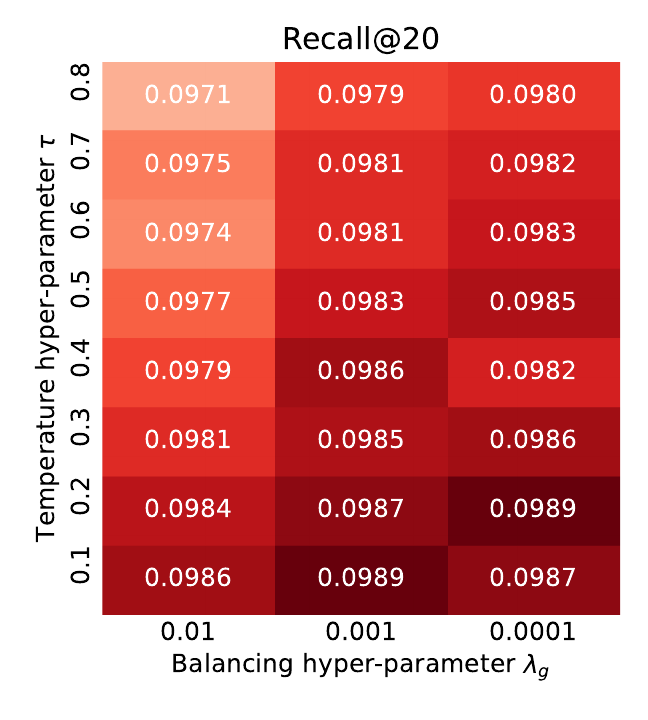}
        }
        \vskip -0.2in
    \caption{Performance of MENTOR with respect to different hyper-parameter pairs ($p$,$\lambda_f$) and  ($\tau$,$\lambda_g$). Darker color denotes better performance of recommendation.}   
    \label{Metrics based on normal}  
\vskip -0.2in
\end{figure}

\subsection{Hyper-parameter Analysis (RQ4)}
\subsubsection{The balancing hyper-parameter $\lambda_{align}$}
The balancing hyper-parameter $\lambda_{align}$ in multilevel cross-modal alignment component is varied in \{1,2,3\}. Fig.~\ref{fig:recall} and Fig.~\ref{fig:ndcg} show the performance trends of MENTOR with different settings of $\lambda_{align}$. We observe that the optimal $\lambda_{align}$ on Baby and Sports datasets is 1, while the optimal $\lambda_{align}$ on Clothing datasets is 2. 

\subsubsection{The pair of hyper-parameters $p$ and $\lambda_{f}$}
We vary the dropout ratio $p$ of MENTOR from 0.1 to 0.7 with a step of 0.1, and vary the balancing parameter $\lambda_{f}$ in \{0.5, 1, 1.5, 2, 2.5\}. The results on three datasets presented in Fig.~\ref{fig:heatmapBaby}-\ref{fig:heatmapClothing}, we find that hyper-parameters $\lambda_{f}$ and $p$ influence each other, so we need to select them in pairs. For Baby and Sports datasets, the optimal values of ($\lambda_{f}$, $p$) pair are (1.5, 0.5) and (1.5, 0.4). Moreover, for Clothing dataset, the optimal value of ($\lambda_{f}$, $p$) pair is (1, 0.3). A possible reason for these results is that the Clothing dataset is more sparse than the other two datasets, which makes it more sensitive to dropout operations.

\subsubsection{The pair of hyper-parameters $\lambda_{g}$ and $\tau$}
The balancing hyper-parameter $\lambda_{g}$ and the temperature hyper-parameter $\tau$ jointly control the feature masking task of the general feature enhancement component. We tune the $\lambda_{g}$ from \{1e-2,1e-3,1e-4\}, and $\tau$ from \{0.1, 0.2, 0.3, 0.4, 0.5, 0.6, 0.7, 0.8\}. From the results on three datasets, Fig.~\ref{fig:heatmapBaby2}-\ref{fig:heatmapClothing2} shows that the best performances are achieved with $\lambda_{g}$ = 1e-3 and $\tau$ = 0.2 on Baby and Sports datasets. On Clothing dataset, the optimal result is achieved with both $\lambda_{g}$ = 1e-3, $\tau$ = 0.1 and $\lambda_{g}$ = 1e-4, $\tau$ = 0.2.

\section{Conclusion}
In this paper, we propose a novel self-supervised learning framework, named MENTOR, for multimodal recommendation. MENTOR introduces a multilevel cross-modal alignment task to align different modalities on data distribution while retaining historical interaction information. Moreover, MENTOR devises a general feature enhancement contrastive learning task from both feature and graph perspectives to improve the model robustness. Our extensive experimental results on several widely used datasets show that MENTOR achieves significant accuracy improvement compared with the state-of-the-art multimodal recommendation methods. In future work, we aim to further reduce the noise induced by different modal information in self-supervised tasks through the large language model (LLM).

\input{sample-sigconf.bbl}


\end{document}